# Impurity Effects on the Superconducting Transition Temperature of $Na_zCoO_2 \cdot yH_2O$

M. Yokoi, H. Watanabe, Y. Mori, T. Moyoshi, Y. Kobayashi and M. Sato

*Department of Physics, Division of Material Science, Nagoya University, Furo-cho, Chikusa-ku Nagoya, 464-8602 Japan*

Superconducting transition temperature $T_c$ of $Na_zCo_{1-x}M_xO_2 \cdot yH_2O$ has been studied for M=Ir and Ga, which can substitute for Co. It has been found that $T_c$ is suppressed by the M-atom doping. The decreasing rates $|dT_c/dx|$ are of the order of 1 K/% and too small to be explained by the pair breaking mechanism for the anisotropic order parameters by non-magnetic impurities. Brief arguments are given on possible origins of the $T_c$ suppression.

corresponding author: M. Sato (e-mail: e43247a@nucc.cc.nagoya-u.ac.jp)

## Introduction

The occurrence of the superconductivity found in the layered cobalt oxide $Na_zCoO_2 \cdot yH_2O$ ($z$~0.3 and $y$~1.3) by Takada *et al.*[1] has attracted much attention, because it has the triangular lattice of Co atoms, in which the electrons are considered to be strongly correlated and frustrated. The system can be obtained from $Na_z \cdot CoO_2$ by de-intercalating $Na^+$ ions and then, by intercalating $H_2O$ molecules. The transition temperature $T_c$ is ~4.5 K. Up to now, various kinds of studies have been carried out to investigate the mechanism of the superconductivity.[1-23] From the experimental side, results of the specific heat ($C$) measurements have been reported by several groups,[6,10,11,13,15] and the nuclear spin-lattice relaxation rate $(1/T_1)$[7,12,16] and the NMR Knight shift ($K$)[8] have also been reported as functions of temperature $T$, for example. Among these works, the exotic nature of the superconductivity is presumed by several groups.

From the theoretical side, a possible role of the interband excitations has been pointed out in the Cooper pair formation in ref. 23. The resemblance of the electronic nature with possible strong fluctuations of this frustrated lattice to that of high-$T_c$ oxides is emphasized by several authors[17,20], and a possibility of the triplet pairing is also conceived by other groups.[18,19,21] In the previous report,[7] the present authors have shown by $^{59}$Co-NQR studies of the system, that the coherence peak of the $^{59}$Co-nuclear spin relaxation rate $1/T_1$ observed by NQR exists just below $T_c$ even though it is much smaller than that expected for ordinary $s$-wave superconductors. We have also reported[8] the NMR Knight shift observed for a peak of the NMR spectra, corresponding to the crystallites oriented with their $y$-axes within the $c$-planes being parallel to the external magnetic field $H$. Based on these results, we have pointed out that the superconducting pairs can probably be considered to be in the singlet state, unless we do not consider a situation where spins of triplet pairs are pinned along the direction perpendicular to the $y$-axis. We have also pointed out that the symmetry may be $s$-like or fully gapped one and added that other kinds of symmetry, $(d+id)$-symmetry, for example, cannot be excluded. Even if the symmetry is $s$-like, it does not necessarily indicate that the superconductivity is not exotic, because the $s$-like symmetry or the fully gapped state may be realized by the pairing mediated by the interband excitations of the electrons.[23]

Another possible way to study the pair symmetry is to examine an effect of the impurity-doping on the superconducting transition temperature of the system: The pair breaking effect of nonmagnetic impurities is negligible in ordinary $s$-wave superconductors, while it can be large as in the case of magnetic impurities in usual $s$-wave superconductors, if the symmetry of the superconducting order parameter has nonzero angular momentum $l$ as the $p$- or $d$-wave one. For nonzero $l$, the absolute initial slope $|dT_c/dx|$ is proportional to the scattering rate $1/\tau$ of the electrons by the impurities. (The relation is described by the pair breaking theory.)[24] Then, if impurities are in the path of the conducting electrons, $|dT_c/dx|$ is expected to be very large. Because this $T_c$-suppression is expected essentially not to be observed, as stated above, for non-magnetic impurities in $s$-wave superconductors, we may have a good way to distinguish if the symmetry of the superconducting order parameter has non-zero $l$ or not.

## Experiments

Polycrystalline samples of $Na_{z'}Co_{1-x}M_xO_2$ (M=Ir and Ga) were prepared by solid reactions at 750 ~850 °C for 24~36 hours, where the nominal value $z'$ was 0.7 or 0.75, depending on series of samples. All samples of $Na_{z'}Co_{1-x}M_xO_2$ (M=Ir and Ga) thus obtained were found to have the $CdI_2$ structure ($P6_3/mmc$) by X-ray powder diffraction with Fe$K$á radiation, where any impurity phases have not been observed for the small values of the nominal concentration $x$ studied here. The lattice parameter $c$ determined by the X-ray studies changed systematically with $x$ for M=Ir, which guarantees that the substitution is successfully carried out. For M=Ga, the lattice parameter $c$ exhibits rather large scattering as a function of $x$. However, because it shows, roughly speaking, a decreasing tendency with increasing $x$ and because the resistivity $\rho$ and the transition temperature $T_c$ of the system exhibit, as shown later, systematic $x$ dependence, it can be considered that the substitution is carried out. (In the actual experiments, besides M=Ir and Ga, several atomic elements such as Cu and Al, for example, were tried to substitute for Co. But, we could not find a firm experimental evidence that these elements really substitute for Co: The lattice parameter did not show systematic dependence on the nominal value $x$. The absolute value and the $T$-dependence of the resistivity $\rho$ did not change systematically with $x$, either. (The superconduting transition temperature was almost independent of the nominal values of $x$.) Then, the obtained data of the doping dependence of $T_c$ on $x$ are, we think, reliable only for M=Ir and Ga. (For M=Rh, the data may also be reliable.) and here these data are presented.

We employed a "rapid heat-up" technique to avoid the Na evaporation.[25] For each sample series obtained through a single series of the processes, the $c$ values were determined by using 5-6 lines between the scattering angles of 20-90º, and the results are shown in Figs. 1 and 5 for M=Ir and Fig. 9 for Ga. The $c$ values at $x$=0 are found to be different among these series probably due to the difference of the real Na concentration, which depends on the preparation conditions such as the sintering temperature $T_a$. (The lattice parameter $c$ decreases with the Na concentration.[25]) For each series of samples, $c$ increases with increasing $x$. The lattice parameter $a$ is rather insensitive to $x$. The resistivity ρ of $Na_{z'}Co_{1-x}M_xO_2$ was measured by the four terminal method. These samples were dried at 150~800 °C before the measurements.

$Na_zCo_{1-x}M_xO_2 \cdot yH_2O$ was prepared from the $Na_{z'}Co_{1-x}M_xO_2$ as follows. Na atoms are de-intercalated by dipping the $Na_{z'}Co_{1-x}M_xO_2$ sample into acetonitrile, in which $Br^-$ ions are dissolved. Then, the sample was washed by distilled water. In this process, $H_2O$ molecules are intercalated. The magnetic susceptibilities χ of the samples were measured by SQUID magnetometer with the magnetic field $H$ of 5 G in the conditions of the zero-field-cooling (ZFC).

**Results and Discussion**

Here, we present results of the studies on two series of Ir-doped system and a series of Ga-doped system. For each series, experimental data of the lattice parameters $c$ and the electrical resistivities ρ of $Na_{z'}Co_{1-x}M_xO_2$, are shown. The magnetic susceptibilities χ due to the shielding diamagnetism and the superconducting transition temperatures $T_c$ of $Na_zCo_{1-x}M_xO_2 \cdot yH_2O$ derived from these series are also shown.

(1) *Series A of Ir-doped system*

For this series, the nominal value of $z'$ is 0.75. Figures 1 and 2 show the lattice parameters $c$ and the electrical resistivities ρ of $Na_{z'}Co_{1-x}Ir_xO_2$, respectively. As stated above, the lattice parameter exhibits the systematic $x$-dependence, indicating that the Ir-substitution for Co is really carried out. The electrical resistivity ρ also exhibits the systematic $x$-dependence and it becomes insulating at $x$ as small as 0.02. The diamagnetic susceptibilities χ and the superconducting transition temperatures $T_c$ of $Na_zCo_{1-x}Ir_xO_2 \cdot yH_2O$ obtained by intercalating $H_2O$ to the samples of $Na_{z'}Co_{1-x}Ir_xO_2$ are shown in Figs. 3 and 4, respectively. The values of $T_c$ are determined by using the observed χ-$T$ curves as the temperatures, at which the absolute values of χ exceeds a certain value with decreasing $T$. (The parentheses in Fig. 4 indicate large ambiguities of the data points.) The $x$-dependence seems to be linear and the slope $|dT_c/dx|$ is estimated to be ~1.0 K/%.

(2) Series B of *Ir-doped system*

For this series, the nominal value of $z'$ is 0.75, too. Figures 5-8 show the data taken for the series B of Ir-doped system similar to those in Figs. 1-4, respectively. The variations of $c$ and ρ with $x$ are, roughly speaking, systematic. For this series of samples, the shielding diamagnetisms are smaller than those observed for the series A. The value of $T_c$ at $x$=0 is larger than that of the series A. However, the initial slope $|dT_c/dx|$ is almost equal to the value observed for the series A. Then, we think that even though the magnitudes of the shielding diamagnetism are different between two series of A and B, we are observing intrinsic behavior of the $T_c$-$x$ curve.

(3) *Series of Ga-doped system*

For this series, the nominal value of $z'$ is 0.7. Results are shown in Figs.9-12, as in the cases of the Ir-doping. Although the increase of $d\rho/dT$ with increasing $x$ found in Fig.10 may not be explained by the simple Matthiessen's law, we have found almost systematic $x$-dependences of $\rho$ and $T_c$. Then, the substitution of Ga for Co is, we think, really carried out. In this system, the slope $|dT_c/dx|$ is ~2.0 K/%, which is about two times larger than the values obtained for the two series of $Na_zCo_{1-x}Ir_xO_2{\cdot}yH_2O$.

We have so far presented the experimental data for three series of $Na_{z'}Co_{1-x}M_xO_2$ and $Na_zCo_{1-x}M_xO_2{\cdot}yH_2O$ (M=Ir and Ga). Because $T_c$ has been found to be, roughly speaking, linear in $x$ in the region of small $x$, the pair breaking theory, which can be described by

$$\ln\{T_c(0)/T_c(x)\}=\psi\{1/2+\hbar/(2\eta\pi k_BT_c(x)\tau)\}-\psi(1/2)$$
$$\equiv \psi(1/2+\alpha/2t)-\psi(1/2)$$

might be applicable to the present systems, where the pair breaking parameter $\alpha\equiv\hbar/\{\eta\pi k_BT_c(0)\tau\}$, $t\equiv T_c(x)/T_c(0)$, $1/\tau$ is the scattering rate by the impurities and $\psi(x)$ is the digamma function. (By using the electron mean free path $\Lambda\sim v_F\tau$ and the coherence length $\xi$ ($=\hbar v_F/\pi\Delta$ in the BCS theory), $\alpha$ is written as $1.76\xi/\Lambda\eta$, where $v_F$ and $\Delta\{=1.76k_BT_c(0)\}$ are the Fermi velocity and the superconducting gap at $x$=$T$=0, respectively.) The $\eta$ value is equal to unity for the spin scattering by magnetic impurities in $s$-wave superconductors and it is equal to 2 for both the potential and spin scatterings by impurities when the order parameter has nonzero angular momentum $l$ (If $l$=0, $\alpha$=0 for non-magnetic impurities and $T_c$ is not suppressed.).

Figure 13 shows the magnetic susceptibilities $\chi$ of the samples of $Na_{z'}Co_{1-x}Ir_xO_2$ (series B) with different $x$. Because the data do not indicate the increase of the number of magnetic moments with increasing $x$, the impurities can be considered to be non-magnetic, which is naturally expected, because non-magnetic Co atoms are substituted with the isoelectronic Ir atoms or with Ga atoms which have only $s$ and $p$ outer electrons. If the superconducting gap parameter has the $s$-wave symmetry, $T_c$ should not, in an ideal case, be suppressed by the present doping.

It is interesting to compare the values of $|dT_c/dx|$ with those of other systems. In Fig.

14, we summarize the $T_c$-$x$ curves obtained in the present experiments together with those of well known systems, $La_{1.85}Sr_{0.15}Cu_{1-x}Zn_xO_4$[26] with the $d$-wave order parameter and $Mg(B_{1-x}C_x)_2$ with the $s$-wave one.[27] The slopes $|dT_c/dx|$ observed for the present sample series have the values similar to that of the $Mg(B_{1-x}C_x)_2$ system and much smaller than that of $La_{1.85}Sr_{0.15}Cu_{1-x}Zn_xO_4$. It may be also useful to note that the values of $|dT_c/dx|$ observed for the spinel system $Cu(Co_{1-x}M_x)_2S_4$ are ~0.9 and ~0.6, for M=Rh and Ir, respectively.[28] The system is considered to have the $s$-wave order parameter from the existence of a significant coherence peak in the $T$-dependence of the nuclear relaxation rate divided by $T$, $1/T_1T$.[29] It has the three dimensional network of the corner-sharing $(Co_{1-x}M_x)_4$ and can be considered to be a three dimensional version of the triangular lattice. For this system, $|dT_c/dx|$ is more than a half of that of $Na_zCo_{1-x}Ir_xO_2{\cdot}yH_2O$. From these results, it is tempting to consider that the present system has the $s$-wave order parameter and that the linear relationship between $T_c$ and $x$ is just caused by a mechanism different from the pair breaking due to the scattering by non-magnetic impurities.

On the other hand, within the framework of the pair breaking theory, the relations $\{T_c(0)-T_c(x)\} \sim\pi\hbar/4\eta k_B\tau \sim \pi^2 T_c(0)\alpha(x)/4 \sim 0.44\pi^2(\xi/\Lambda\eta)\times T_c(0)$ hold in the region of small $x$ (Note that $\xi\times T_c(0)$ does not depend on $T_c(0)$. ). To explain the initial slope $|dT_c/dx| \sim 1$ K/%, $\xi/(\Lambda\eta) = \hbar v_F/\{1.76\pi k_B T_c(0)\}/(v_F\tau)$ should be of the order of ~1/20. If $\xi$ is chosen to be ~100 Å,[10] $\Lambda$ has to be as large as 1000 Å at $x$=0.01, which seems to be too large as compared with the average separation of neighboring impurities (~30 Å). The much smaller values of $|dT_c/dx|$ for the present systems than that of $La_{1.85}Sr_{0.15}Cu_{1-x}Zn_xO_4$ do not seem to be easily understood only by the naive consideration of the pair breaking by non-magnetic impurities, either

Then, what is the primary mechanism of the observed $T_c$ suppression? It is, roughly speaking, to be linear in $x$, but the rate $|dT_c/dx|$ seems to be too small to understand by the pair breaking mechanism. We may have to ask why $Cu(Co_{1-x}M_x)_2S_4$ with the $s$-symmetry of the order parameter exhibits the linear $T_c$-decrease in $x$,[28] because the similar origin of the $T_c$ suppression may be relevant to the present case. For example, effects of the carrier localization on $T_c$ may be considered as one of possible origins, because for M=Ir, the relatively small amount of the doping to the host material $Na_{z'}CoO_2$ seems to induce the upturn of the resistivity with decreasing $T$ (see Figs. 2 and 6).

Before summarizing the results of the present study, we also add followings.[30] The Knight shifts $K_y$ studied as functions of $T$ at several fixed $H$ values (1.5 T $\leq H \leq$ 4.5 T) indicate that the spin susceptibility decreases with decreasing $T$ below $T_c$. The $H$- and $T$-dependences of $K_y$ cannot be explained by considering the superconducting diamagnetism, as reported in our previous paper.[8)] The magnetic field $H_{c2}$ determined from the $T$ dependence of $K_y$ reproduces the $H_{c2}$-$T$ curve for $\boldsymbol{H}$ applied within the $c$ plane reported in ref. 10. The $H_{c2}$ value at $T \rightarrow 0$ can be understood by the effect of the Zeeman splitting on singlet pairs, indicating again that the superconducting pairs are definitely in the singlet state. It gives a restriction that the pairs have even parity.

In order to fully understand the observed values of $|dT_c/dx|$, we have to examine various quantities and study the electronic state of the present system in detail. In summary, we have presented the results of the studies on the impurity-doping effects on $T_c$ of $Na_zCoO_2 \cdot yH_2O$ and shown that the values of $|dT_c/dx|$ are ~1 K/%, which is too small to be explained by the pair breaking mechanism for the anisotropic ($l \neq 0$) order parameters by non-magnetic impurities.

Acknowledgments- The work is supported by Grants-in-Aid for Scientific Research from the Japan Society for the Promotion of Science (JSPS) and by Grants-in-Aid on priority area from the Ministry of Education, Culture, Sports, Science and Technology.

References

1. K. Takada, H. Sakurai, E. Takayama-Muromachi, F. Izumi, R. A. Dilanian and T. Sasaki: Nature **422** (2003) 53.
2. M. L. Foo, R. E. Schaak, V. L. Miller, T. Klimczuk, N. S. Rogado, Y. Wang, G. C. Lau, C. Craley, H. W. Zandbergen, N. P. Ong and R. J. Cava: Solid State Commu. **127** (2003) 33.
3. H. Sakurai, K. Takada, S. Yoshii, T. Sasaki, K. Kindo and E. Takayama-Muromachi: Phys. Rev. B **68** (2003) 132507.
4. B. Lorenz, J. Cmaidalka, R. L. Meng and C. W. Chu: Phys. Rev. B **68** (2003) 132504.
5. R. E. Schaak, T. Klimczuk, M. L. Foo and R. J. Cava: Nature **424** (2003) 527.
6. G. Cao, C. Feng, Y. Xu, W. Lu, J. Shen, M. Fang and Z. Xu: J. Phys.: Condens. Matter **15** (2003) L519.
7. Y. Kobayashi, M. Yokoi and M. Sato: J. Phys. Soc. Jpn. **72** (2003) 2161.
8. Y. Kobayashi, M. Yokoi and M. Sato: J. Phys. Soc. Jpn. **72** (2003) 2453.
9. R. Jin, B.C. Sales, P. Khalifah and D. Mandrus: Phys. Rev. Lett. **91** (2003) 217001.
10. F. C. Chou, J. H. Cho, P. A. Lee, E. T. Abel, K. Matan and Y. S. Lee: cond-mat/0306659.
11. B. G. Ueland, P. Schiffer, R. E. Schaak, M.L. Foo, V. L. Miller and R. J. Cava: cond-mat/0307106.
12. T. Fujimoto, G. Zheng, Y. Kitaoka, R. L. Meng, J. Cmaidalka and C. W. Chu: cond-mat/0307127.
13. H. D. Yang, J. -Y. Lin, C. P. Sun, Y. C. Kang, K. Takada, T. Sasaki, H. Sakurai and E. Takayama-Muromachi: cond-mat/0308031.
14. J. Cmaidalka, A. Baikalov, Y. Y. Xue, R. L. Meng and J. Cmaidalka and C. W. Chu: cond-mat/0308301.
15. B. Lorenz, J. Cmaidalka, R. L. Meng and C. W. Chu: cond-mat/0308143.
16. K. Ishida, Y. Ihara, Y. Maeno, C. Michioka, M. Kato, K. Yoshimura, T. Takada, T. Sasaki, H. Sakurai and E. Takayama-Muromachi: J. Phys. Soc. Jpn. **72** (2003) 3041.
17. G. Baskaran: Phys. Rev. Lett. **91** (2003) 097003.
18. B. Kumar and B. S. Shastry: Phys. Rev. B **68** (2003) 104508.
19. Q.-H. Wang , D. -H. Lee and P. A. Lee: cond-mat/0304377.
20. M. Ogata: J. Phys. Soc. Jpn. **72** (2003) 1839.
21. A. Tanaka and X. Hu: cond-mat/0304409
22. D. J. Singh: Phys. Rev. B **68** (2003) 020503.
23. K. Sano and Y. Ono: J. Phys. Soc. Jpn. **72** (2003) 1847.
24. A. A. Abrikosov and L. P. Gor'kov: JETP **12** (1961) 1243.
25. T. Motohashi, E. Naujalis, R. Ueda, K. Isawa, M. Karppinen and H. Yamauchi: Appl.

Phys. Lett. 79 (2001) 1480.
26. H. Harashina, T. Nishikawa, T. Kiyokura, S. Shamoto, M. Sato and K. Kakurai: Physica C **212** (1993) 142.
27. T. Takenobu, T. Ito, Dam Hieu Chi, K. Prassides and Y. Iwasa: Phys. Rev. B **64** (2001) 134513.
28. N. Aito and M. Sato: in preparation.
29. H. Sugita, S. Wada, K. Miyatani, T. Tanaka and M. Ishikawa; Physica B 284-288 (2000) 473.
30. Y. Kobayashi, M. Yokoi and M. Sato: unpublished data.

Figure captions

Fig. 1 Lattice parameters $c$ of the sample series A of $Na_{z'}Co_{1-x}Ir_xO_2$ at room temperature are plotted against $x$. The nominal value of $z'=0.75$.

Fig. 2 Temperature dependence of the resistivities $\rho$ of the sample series A of $Na_{z'}Co_{1-x}Ir_xO_2$ is shown for various $x$. The nominal value of $z'=0.75$.

Fig. 3 Magnetic susceptibilities $\chi$ due to the shielding diamagnetism for the sample series A of $Na_zCo_{1-x}Ir_xO_2 \cdot yH_2O$ are shown against $x$.

Fig. 4 Superconducting transition temperatures $T_c$ of the sample series A of $Na_zCo_{1-x}Ir_xO_2 \cdot yH_2O$ are shown against $x$. The parentheses indicate possible large error bars.

Fig. 5 Lattice parameters $c$ of the sample series B of $Na_{z'}Co_{1-x}Ir_xO_2$ at room temperature are plotted against $x$. The nominal value of $z'=0.75$.

Fig. 6 Temperature dependence of the resistivities $\rho$ of the sample series B of $Na_{z'}Co_{1-x}Ir_xO_2$ is shown for various $x$. The nominal value of $z'=0.75$.

Fig. 7 Magnetic susceptibilities $\chi$ due to the shielding diamagnetism for the sample series B of $Na_zCo_{1-x}Ir_xO_2 \cdot yH_2O$ are shown against $x$.

Fig. 8 Superconducting transition temperatures $T_c$ of the sample series B of $Na_zCo_{1-x}Ir_xO_2 \cdot yH_2O$ are shown against $x$. The parentheses indicate possible large error bars.

Fig. 9 Lattice parameters $c$ of the sample series of $Na_{z'}Co_{1-x}Ga_xO_2$ at room temperature are plotted against $x$. The nominal value of $z'=0.7$.

Fig. 10 Temperature dependence of the resistivities $\rho$ of the sample series of $Na_{z'}Co_{1-x}Ga_xO_2$ is shown for various $x$. The nominal value of $z'=0.7$.

Fig. 11 Magnetic susceptibilities $\chi$ due to the shielding diamagnetism for the sample series of $Na_zCo_{1-x}Ga_xO_2 \cdot yH_2O$ are shown against $x$.

Fig. 12 Superconducting transition temperatures $T_c$ of the sample series of $Na_zCo_{1-x}Ga_xO_2 \cdot yH_2O$ are shown against $x$. The parentheses indicate possible large error bars.

Fig. 13 The magnetic susceptibilities of $Na_{z'}Co_{1-x}Ir_xO_2$ are shown against $T$ for three Ir concentration $x$. The nominal value of $z'=0.75$.

Fig. 14 Superconducting transition temperatures $T_c$ of various superconducting systems are plotted as functions of the doped atom concentrations.

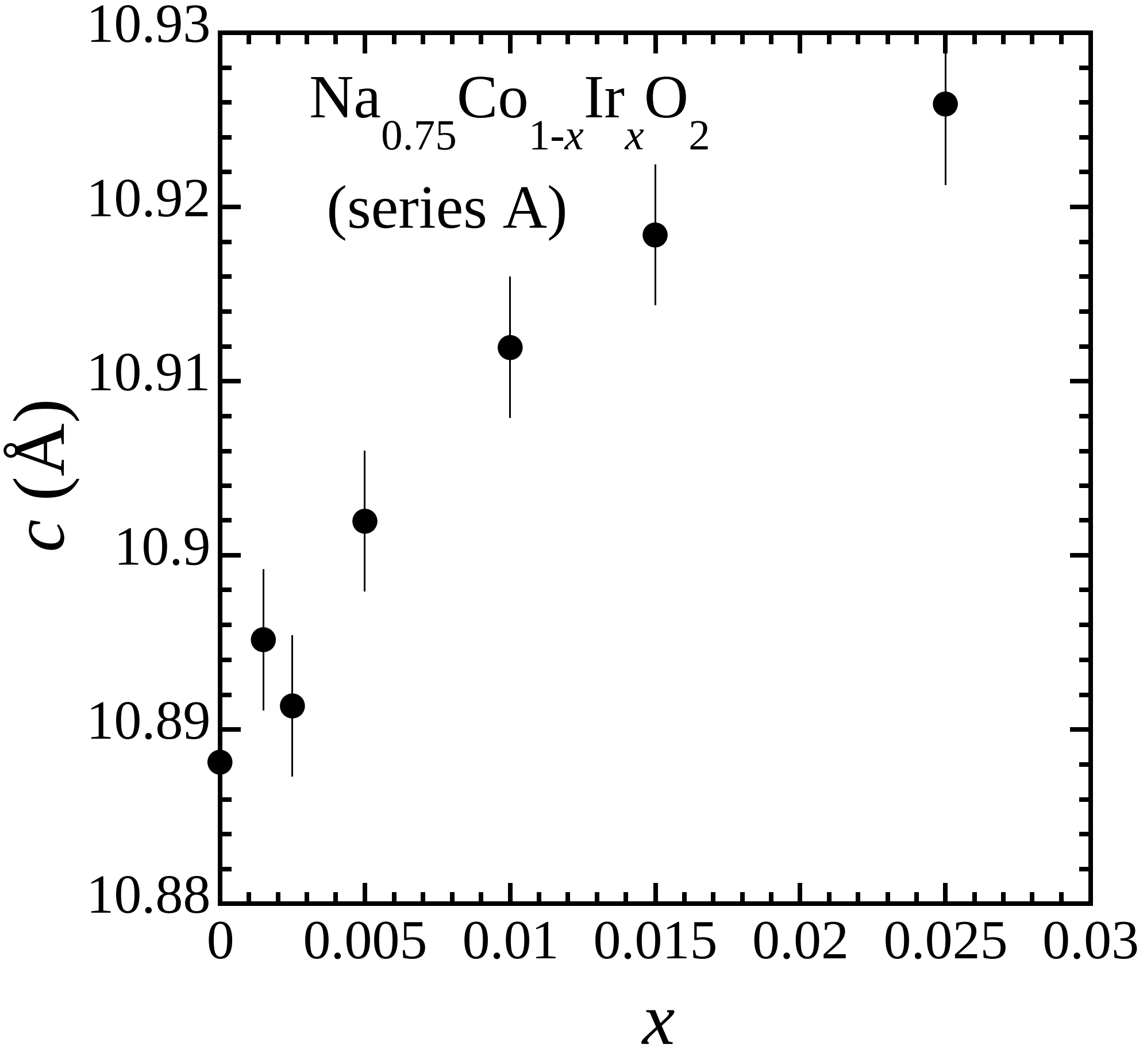

Fig.1

Yokoi *et al*.

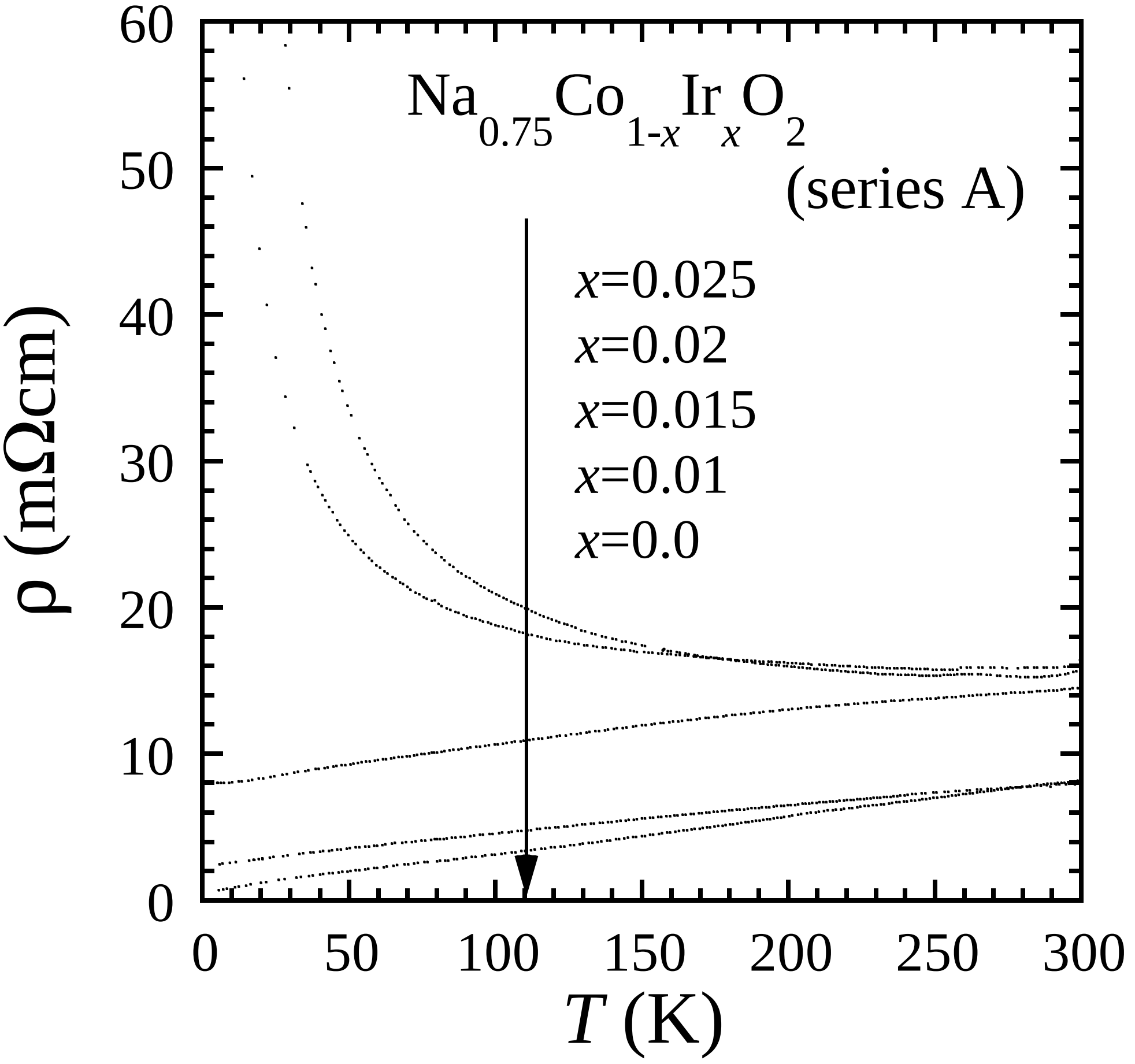

Fig.2
Yokoi *et al.*

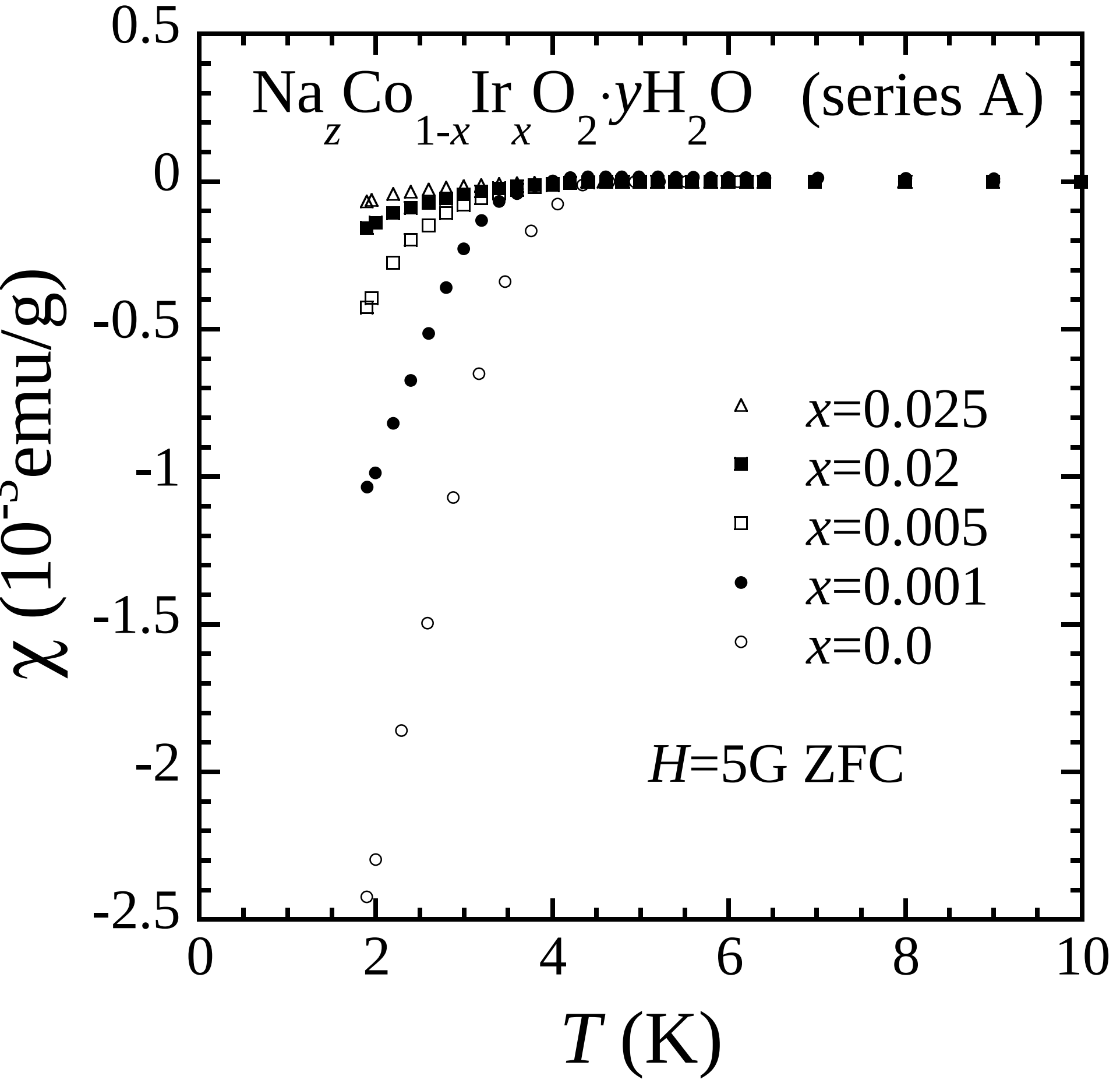

Fig.3
Yokoi *et al*.

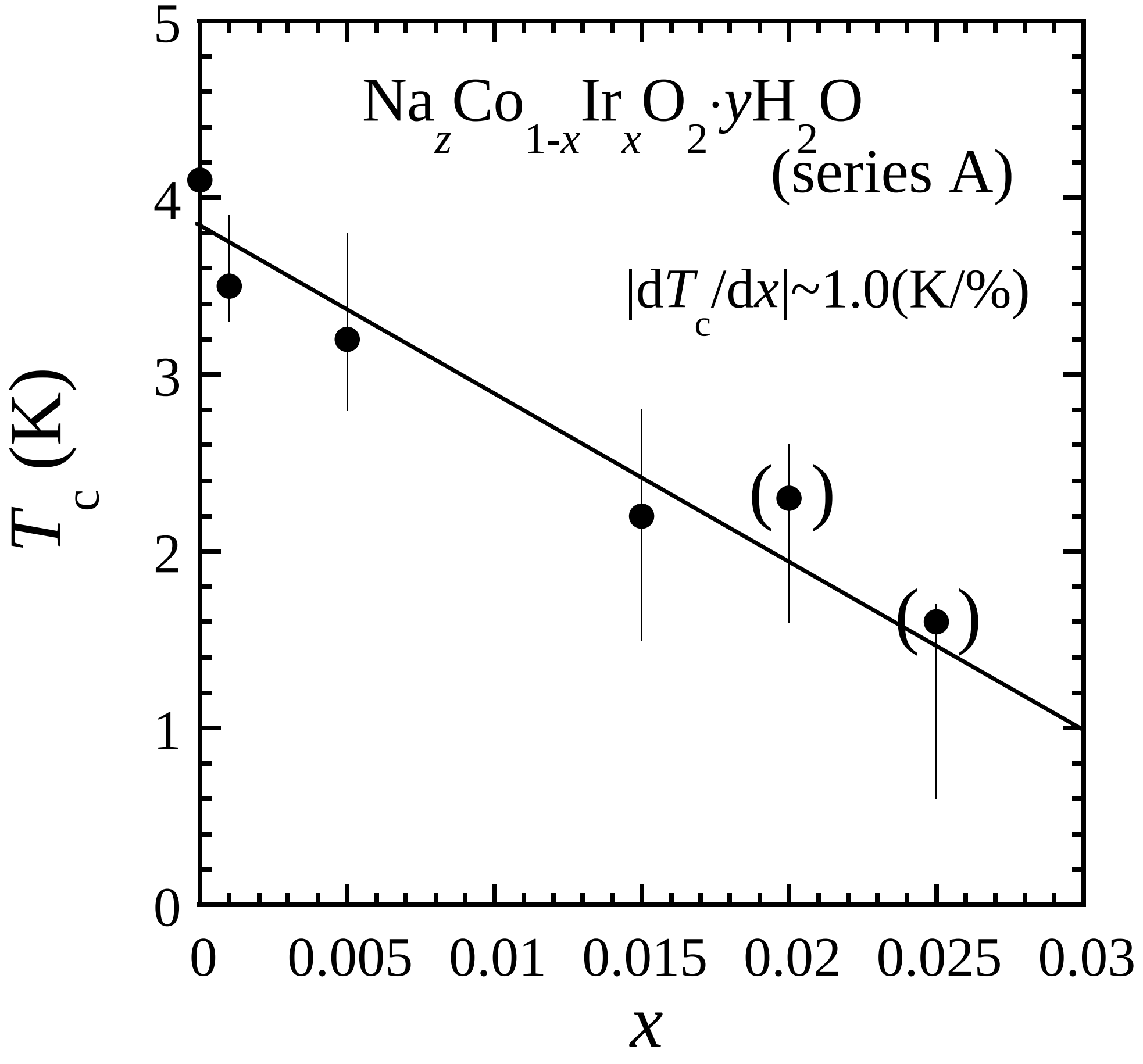

Fig.4
Yokoi *et al*.

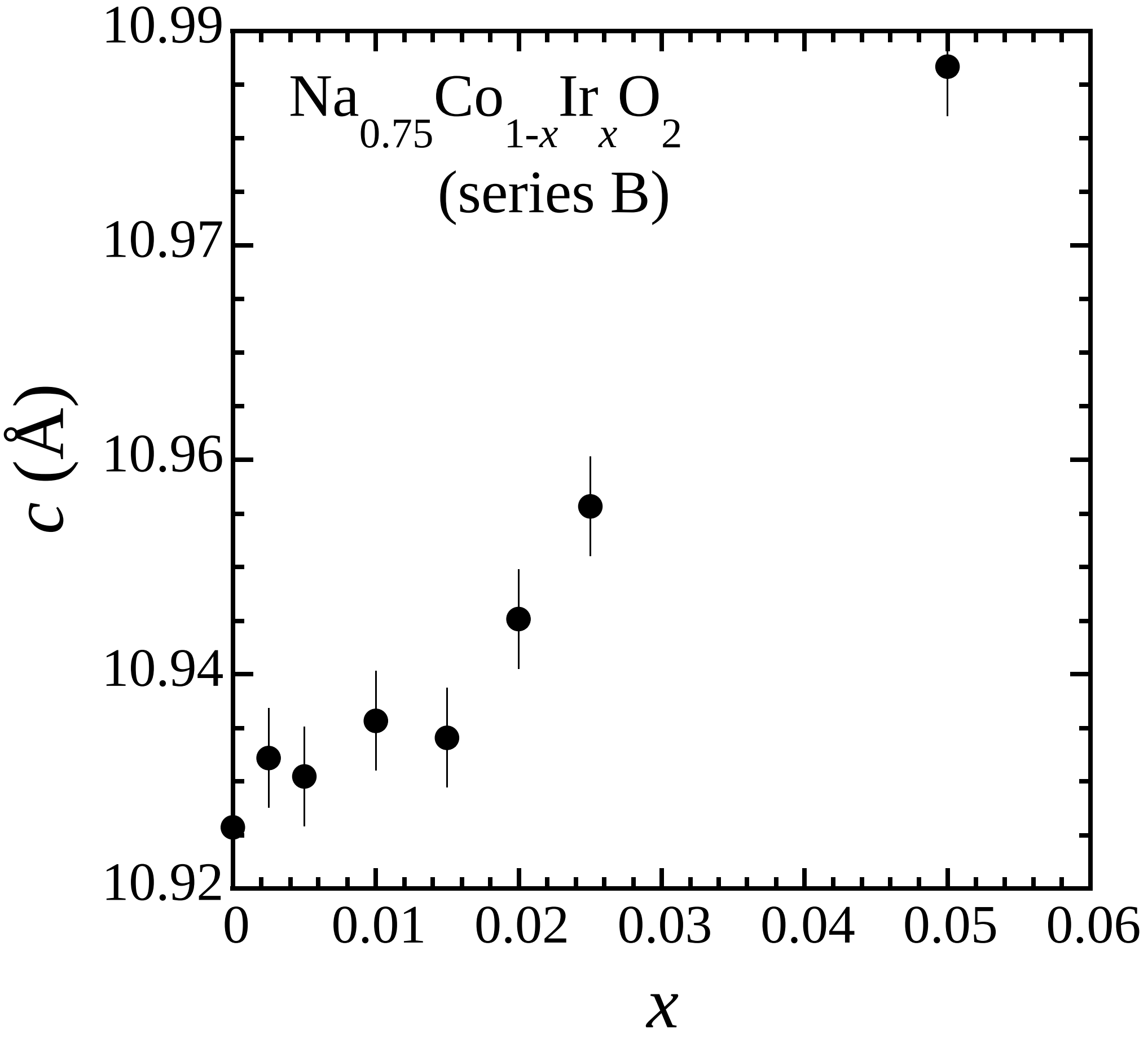

Fig.5
Yokoi *et al*.

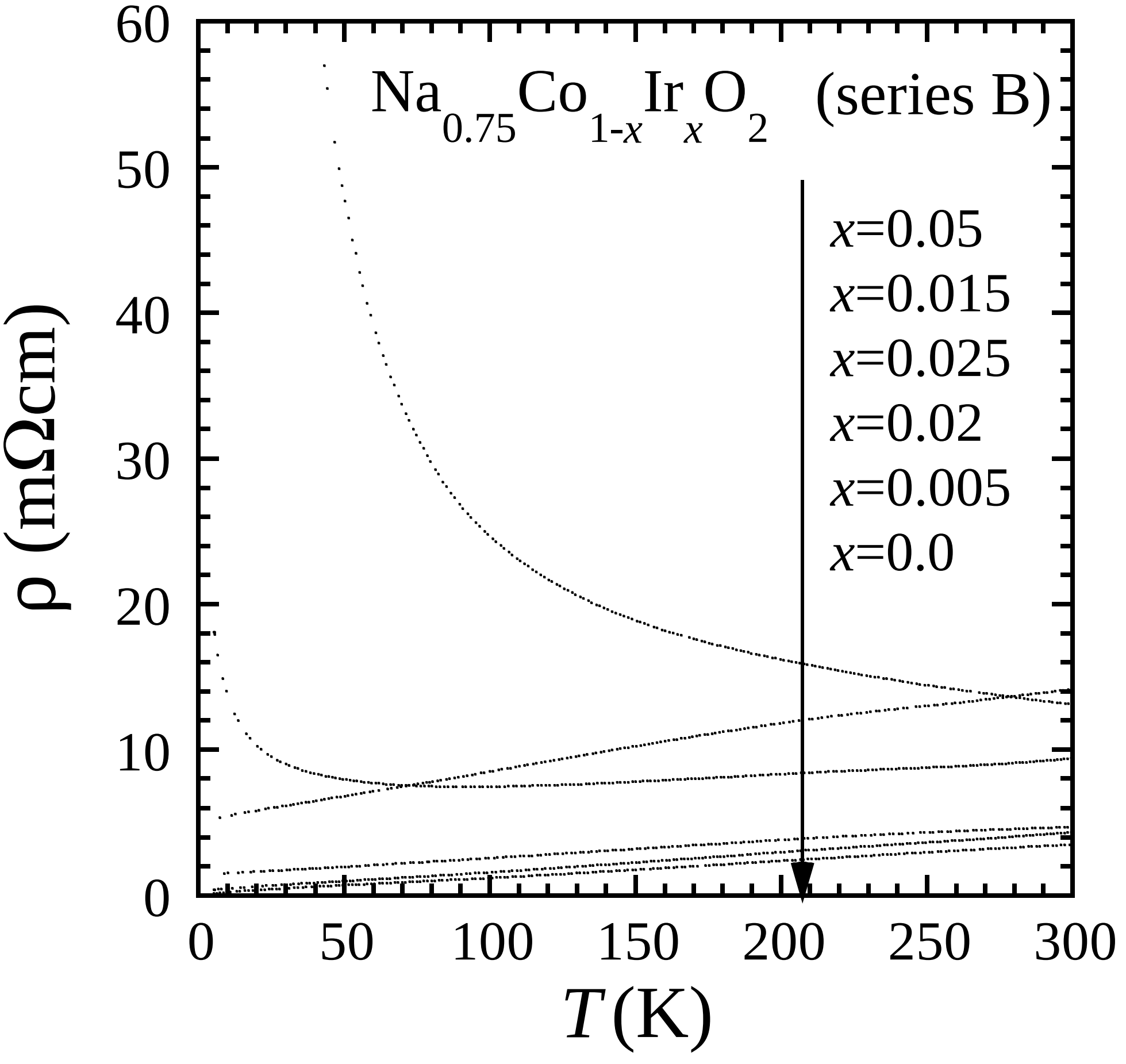

Fig.6
Yokoi *et al*.

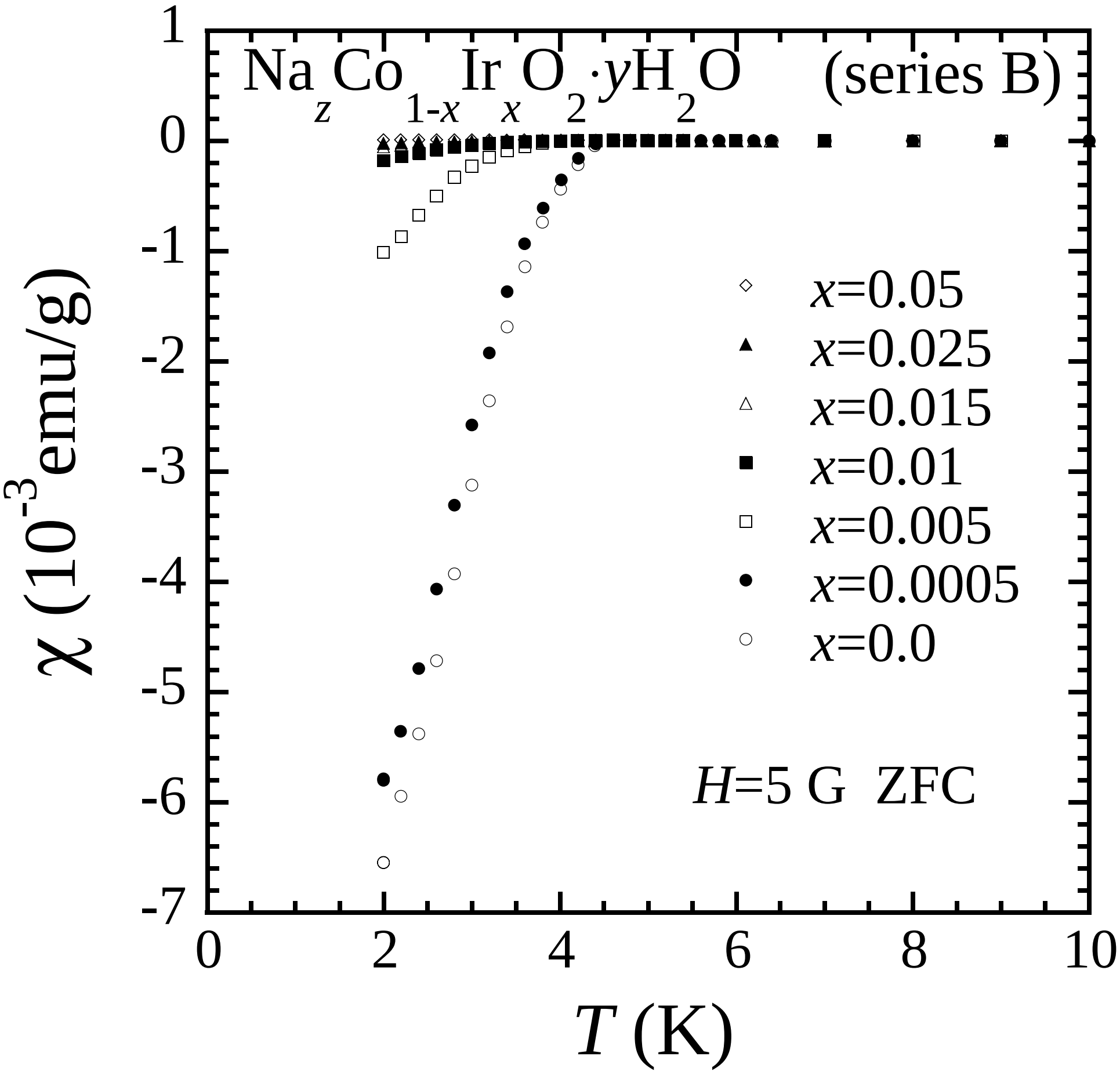

Fig.7
Yokoi *et al*.

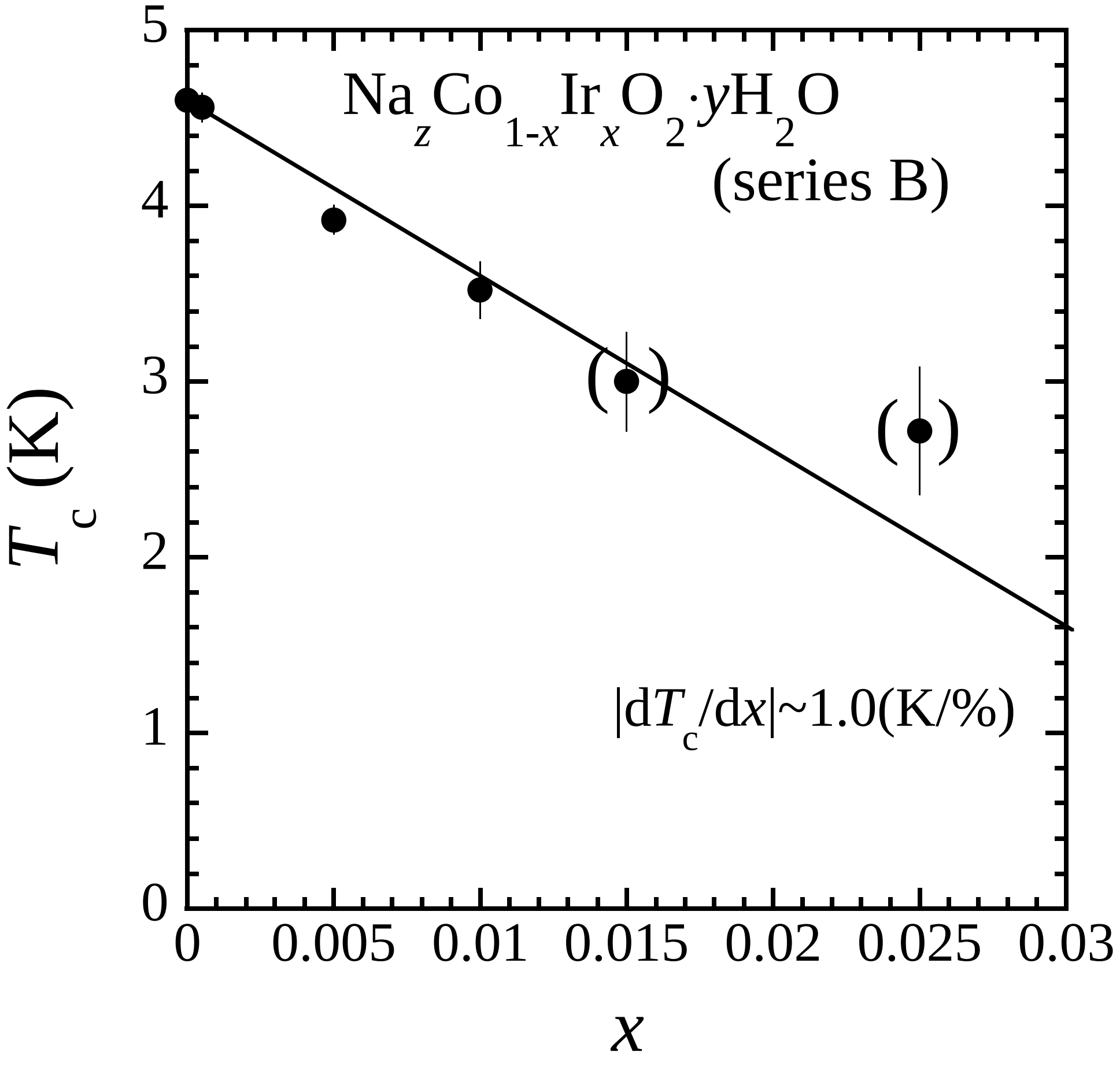

Fig.8
Yokoi *et al*.

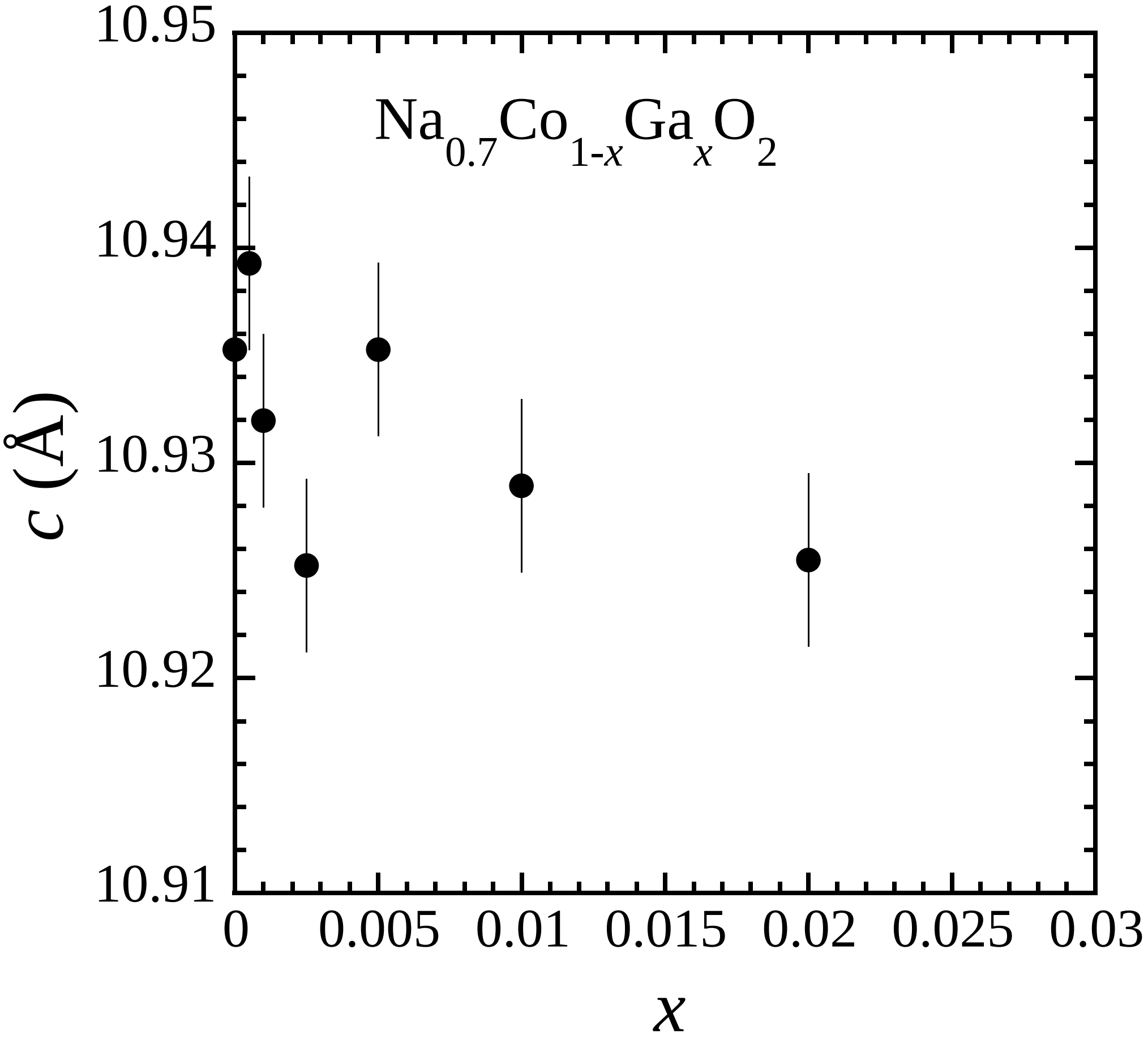

Fig.9
Yokoi *et al*.

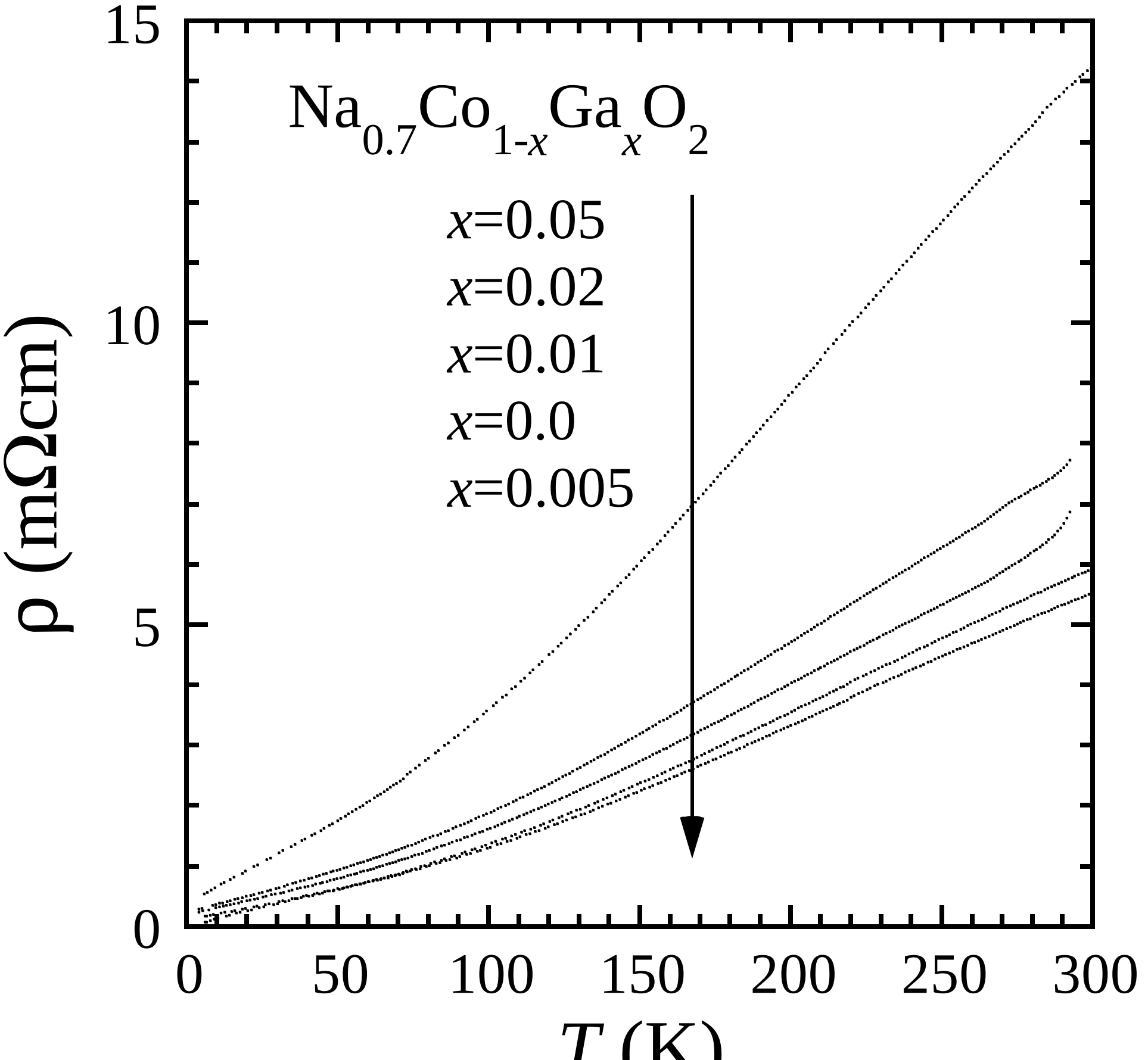

Fig.10
Yokoi *et al.*

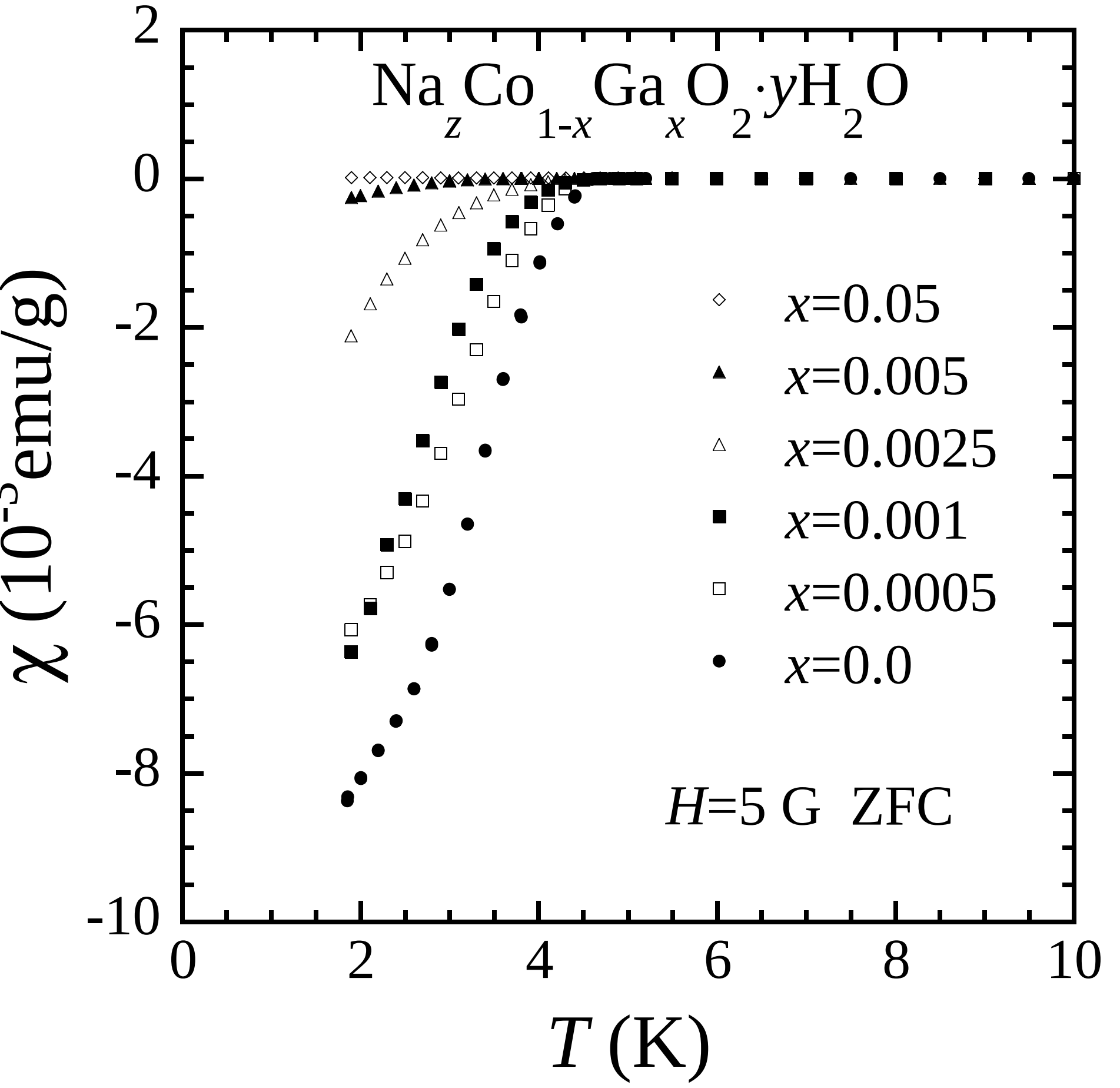

Fig.11
Yokoi *et al.*

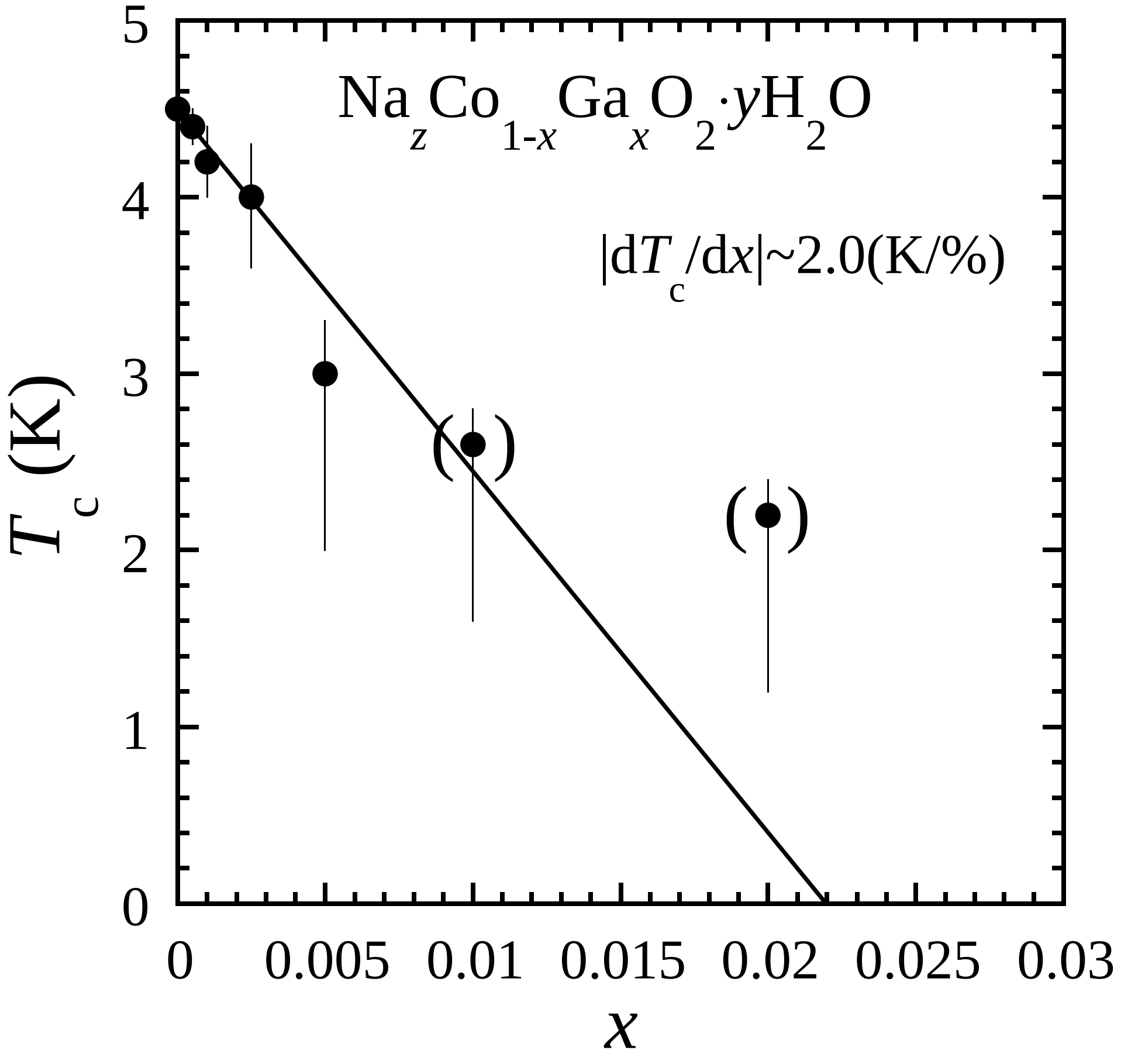

Fig.12
Yokoi *et al*.

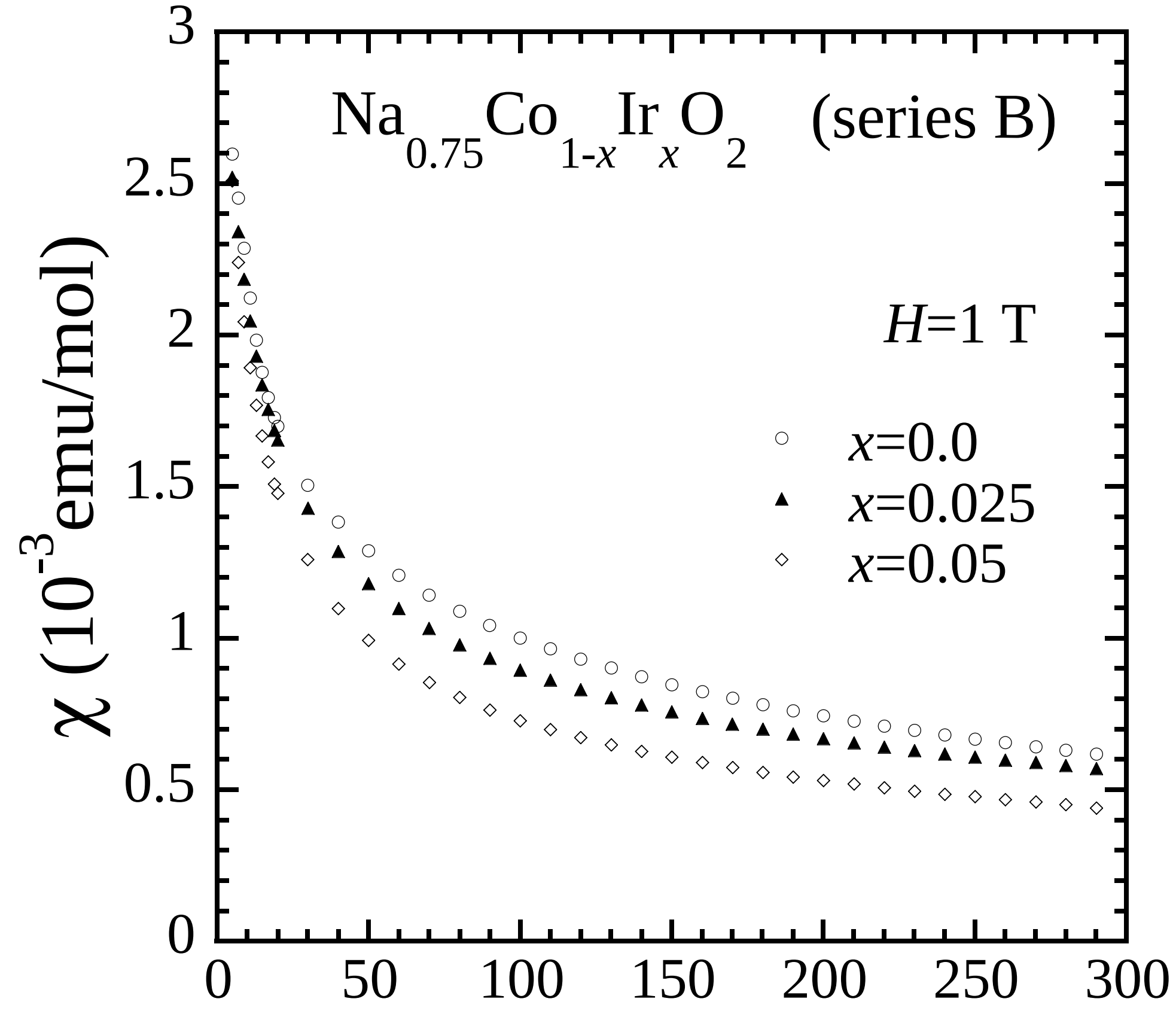

Fig.13
Yokoi *et al*.

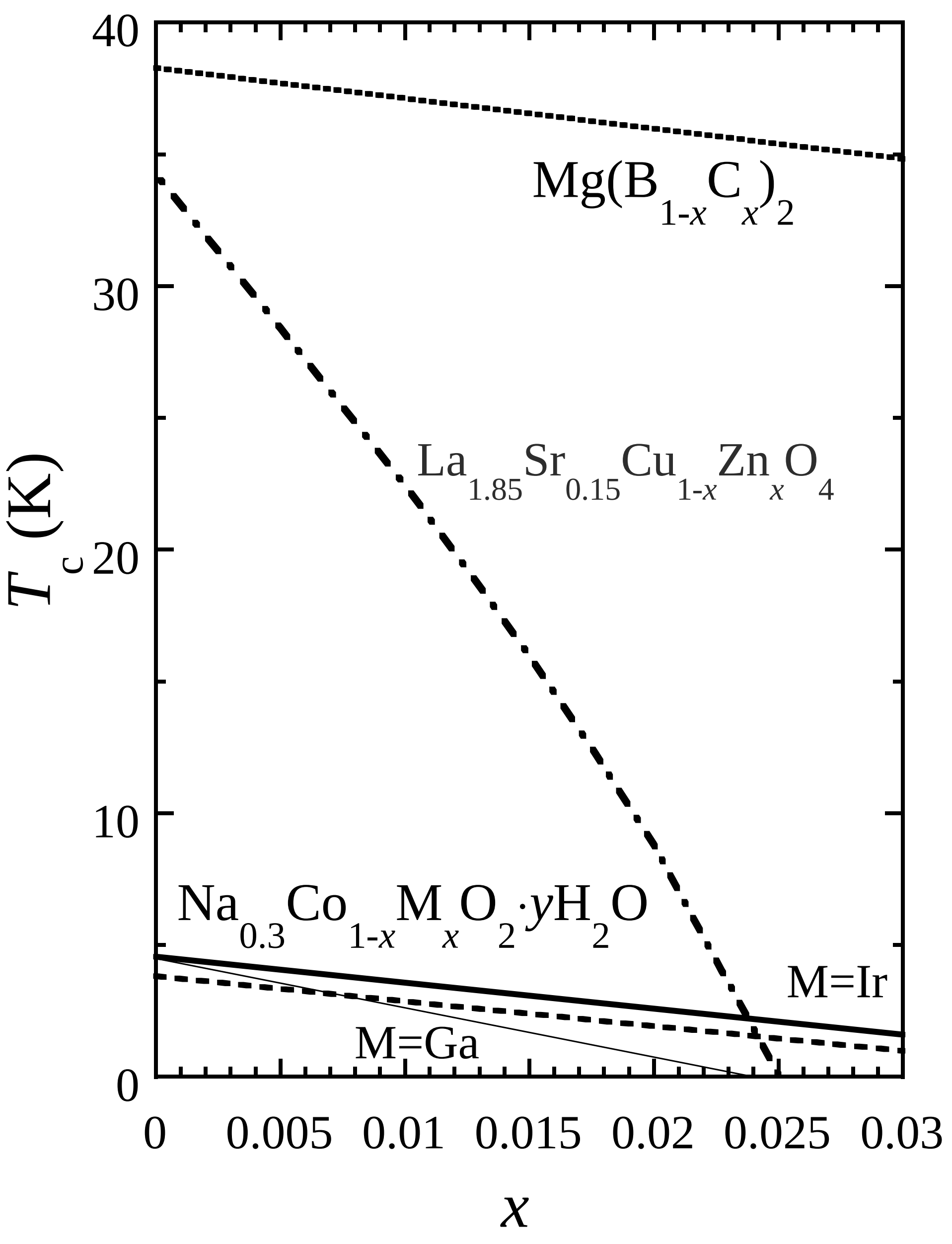

Fig.14
Yokoi *et al.*